\documentstyle{article}
\begin{document}
\begin{titlepage}
\begin{flushright}
Fermilab-Pub-97/317-T
\end{flushright}
\begin{center}
{\Large \bf Measurement of the $W$ Boson Mass at the LHC}\\
\vspace{1cm}
{\large S.~Keller\footnote{keller@fnal.gov} \\ \vspace{0.5cm} 
and \\ \vspace{0.5cm}
J.~Womersley\footnote{womersley@fnal.gov}\\ \vspace{0.5cm} {\it Fermi National
Accelerator Laboratory,\\ P.O. Box 500, Batavia, IL 60510, U.S.A.
\footnote{operated by the Universities Research Association, Inc., for
the U.S. Department of Energy} }}
\end{center}
\begin{abstract}
{We explore the ability of the Large Hadron Collider to
measure the mass of the $W$ boson.  We believe that a precision better
than $\sim 15$~MeV could be attained, based on a year of operation at
low luminosity ($10^{33}\,{\rm cm}^{-2}\,{\rm s}^{-1}$).  If this is
true, this measurement will be the world's best determination of the
$W$ mass.  }
\end{abstract}
\end{titlepage}

\section{Introduction}

The mass of the $W$ boson, $m_W$, is one of the fundamental parameters
of the Standard Model.  As is well known~\cite{wolf}, a precise
measurement of $m_W$, along with other precision electroweak
measurements, will lead, within the Standard Model, to a strong
indirect constraint on the mass of the Higgs boson.  Once the Higgs
itself is found, this will provide a consistency test of the Standard
Model and, perhaps, evidence for physics beyond.  The precise
measurement of $m_W$ is therefore a priority of future colliders.
LEP2 and Run II at Fermilab ($\int\!{\cal L}dt=1$~fb$^{-1}$) are
aiming for an uncertainty on $m_W$ of about 40 MeV~\cite{lep} and 35
MeV~\cite{tev2000}, respectively.  An upgrade of the
Tevatron~\cite{tevah}, beyond Run II, might be possible, with a goal
of an overall integrated luminosity of ${\cal O}(30 \mbox{fb}^{-1})$
and a precision on $m_W$ of about 15 MeV.  Clearly, hadron colliders
have had and will continue to have a significant impact on the
measurement of $m_W$.  In this short paper~\cite{previous} we
investigate the potential to measure $m_W$ at the Large Hadron
Collider (LHC).  The LHC will provide an extremely copious source of
$W$ bosons, thus allowing in principle for a statistically very
precise measurement.

In Section 2 we consider the detector capabilities, in Section 3 the 
theoretical uncertainties, and in Section 4 the experimental uncertainties.  
We present our conclusions in Section 5.  

\section{Detector Capabilities}

A potential problem is that the general-purpose LHC detectors might
not be able to trigger on leptons with sufficiently low transverse
momentum ($p_T$) to record the $W$ sample needed for a measurement of
$m_W$.  While this may be true at the full LHC luminosity
($10^{34}\,{\rm cm}^{-2}\,{\rm s}^{-1}$) it does not appear to be the
case at $10^{33}\,{\rm cm}^{-2}\,{\rm s}^{-1}$.  Based on a full GEANT
simulation of the calorimeter, the CMS isolated electron/photon
trigger~\cite{cmsetrig} should provide an acceptable rate ($<5$~kHz at
level 1) for a threshold setting of $p_T^{e,\gamma} > 15$~GeV/c. This
trigger will be fully efficient for electrons with $p_T^{e} >
20$~GeV/c.  The CMS muon trigger~\cite{cmsmutrig} should also operate
acceptably with a threshold of $p_T^{\mu} > 15-20$~GeV/c at
$10^{33}\,{\rm cm}^{-2}\,{\rm s}^{-1}$.  
ATLAS should have similar capabilities. 
It is likely that the
accelerator will operate for at least a year at this `low' luminosity
to allow for studies which require heavy quark tagging (e.g.,
$B$-physics).  This should provide an integrated luminosity of the
order of $10fb^{-1}$.

The mean number of interactions per crossing, $I_C$, is about 2 at the
low luminosity.  This is actually lower than the number of
interactions per crossing during the most recent run (IB) at the
Fermilab Tevatron.  In this relatively quiet environment it should be
straightforward to reconstruct electron and muon tracks with good
efficiency.  Furthermore, both 
the ATLAS\cite{ATLAS} and CMS\cite{CMS} detectors offer
advances over their counterparts at the Tevatron for lepton
identification and measurement: they have precision electromagnetic
calorimetry (liquid argon and PbWO$_4$ crystals, respectively) and
precision muon measurement (air core toroids and high field solenoid,
respectively).

The missing transverse energy will also be well measured thanks to the
small number of interactions per crossing and the large pseudorapidity
coverage ($|\eta|<5$) of the hadronic calorimeters.  The so-far
standard transverse-mass technique for determining $m_W$ should thus
continue to be applicable.  This is to be contrasted with the problem
that the increase in $I_C$ will create for Run II (and beyond) at the
Tevatron.  In Ref.~\cite{tev2000}, it was shown that it will
substantially degrade the measurement of the missing transverse energy
and therefore the measurement of $m_W$.

\section{Theoretical Uncertainties}

\begin{figure}[th]
\vskip 3in
\includegraphics{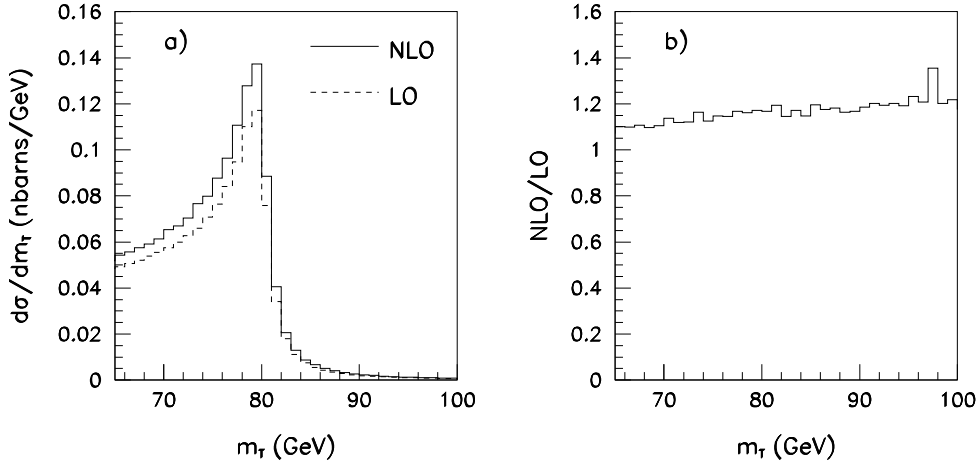}
\caption{a) LO calculation (dashed line) and 
NLO QCD calculation (solid line) of the $m_T$ distribution at the LHC.  
See text for the cuts.  
b) Ratio of the NLO calculation over the LO calculation
as a function of $m_T$.}
\label{fig:lhc}
\end{figure}

\begin{figure}[th]
\vskip 3in
\includegraphics{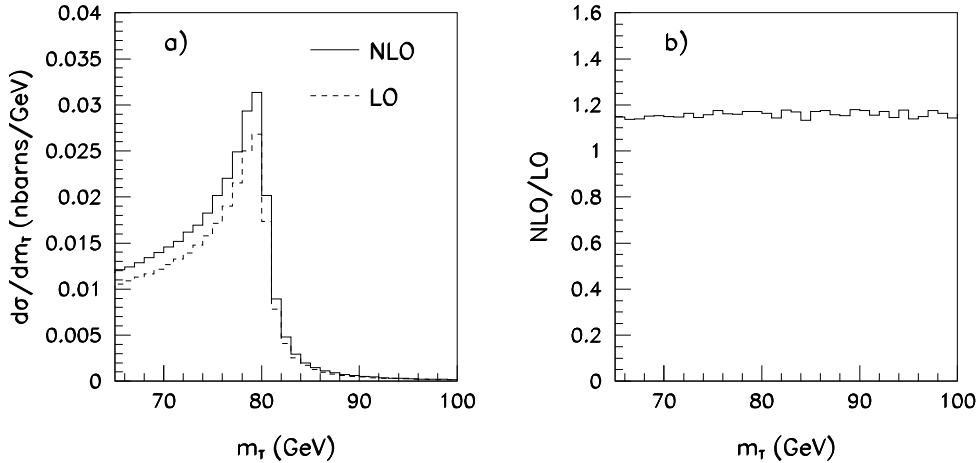}
\caption{Same as in Fig.~\ref{fig:lhc} but for the Tevatron.}
\label{fig:tev}
\end{figure}

Large theoretical uncertainties arising from substantial QCD corrections to
$W$ production at the LHC energy could deteriorate the possible measurement 
of $M_W$.  In Fig.~\ref{fig:lhc}a,
we present the leading order (LO) calculation and next-to-leading
order (NLO) QCD calculation~\cite{walt} of the transverse mass
distribution ($m_T$) at the LHC (14~TeV, $pp$ collider) in the region
of interest for the extraction of the mass.  We used the
MRSA~\cite{mrsa} set of parton distribution functions, and imposed a
charged lepton (electron or muon) rapidity cut of 1.2, as well as a
charged lepton $p_T$ and missing transverse energy cut of 20 GeV.  We
used $m_W$ for the factorization and renormalization scales.  No
smearing effects due to the detector were included in our calculation.
The uncertainty due to the QCD corrections can be gauged by
considering the ratio of the NLO calculation over the LO calculation.
This ratio is presented in Fig.~\ref{fig:lhc}b as a function of
$m_T$.  As can be seen, the corrections are not large and vary between
10\% and 20\%.  For the extraction of $m_W$ from the data, the
important consideration is the change in the shape of the
$m_T$-distribution.  As can be seen from Fig.~\ref{fig:lhc}b, the
corrections to the shape of the $m_T$-distribution are at the 10\%
level.  Note that an increase in the charged lepton $p_T$ cut has the
effect of increasing the size of the shape change (it basically
increases the slope of the NLO over LO ratio), such that for the
theoretical uncertainty is is better to keep that cut as low as
possible.  For comparison, in Fig.~\ref{fig:tev} we present the same
distributions as in Fig.~\ref{fig:lhc} for the Tevatron energy
(1.8~TeV, $p \bar{p}$ collider).
The same cuts as for the LHC were applied.  As can be seen the
corrections are of the order of 20\% and change the shape very little.

Currently, the estimated uncertainty on $m_W$ associated with
modelling the transverse momentum distribution of the W ({\it i.e.}
due to QCD corrections) is of the order of 10~MeV at the
Tevatron\cite{wmassd0}.  On the one hand, the larger QCD corrections at
the LHC suggest that the uncertainty will also be larger.  On the
other hand, the $p_T$ distribution of the W can be constrained by data
(both $W$ and $Z$) and the significant increase in statistics
available, first at the upgraded Tevatron and then at the LHC, should 
keep the uncertainty under control.  Note also that even
though the shape change due to QCD corrections is undoubtedly larger
at the LHC than at the Tevatron, in absolute terms it is still small
and a next-to-next-to leading order calculation might be able to
reduce the theoretical uncertainty to an acceptable level.  Although
such a calculation does not yet exist for the $m_T$-distribution, one
may certainly imagine that it will be before any data become available
at the LHC.

\begin{figure}[th]
\vskip 3in
\includegraphics{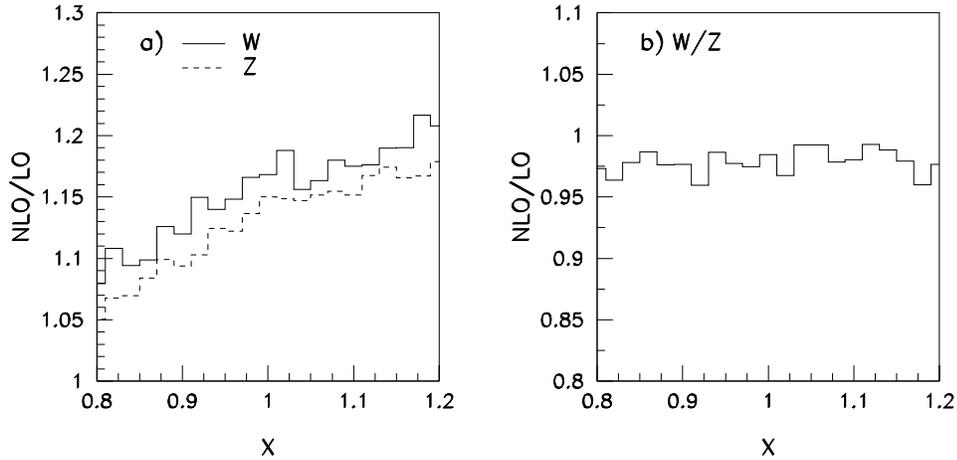}
\caption{a) Ratio of NLO/LO calculations for the distribution of
events as a function of the mass-scaled transverse mass $X$ at the 
LHC energy. See text for the cuts.  b) NLO/LO for the quantity defined as
the ratio of the number of $W$ events to the number of
$Z$ events at a given $X$.  
\label{fig:ratio}}
\end{figure}

An alternative would be to use an observable with yet smaller QCD
corrections.  Recently~\cite{gandk}, it was pointed out that the ratio
of $W$ to $Z$ observables (properly scaled by the respective masses)
are subject to smaller QCD corrections than the observables
themselves.  This is illustrated in Fig.~\ref{fig:ratio} for the
transverse mass.  In Fig.~\ref{fig:ratio}a the ratio of NLO/LO
calculations for the distribution of events as a function of the
mass-scaled transverse mass $X$ is presented; $X = m_T^W/m_W$ for the
$W$ and $m_T^Z/m_Z$ for the $Z$.  The cuts for the $W$ case are as
described before.  The $Z$ is required to have one lepton with $\eta <
1.2$ and the $p_T$ cuts are scaled proportionally to the mass compared
to the $p_T$ cut in the $W$ case~\footnote{This is done to avoid large
corrections to the ratio of $W$ to $Z$ observables close to the cut.}.
Fig.~\ref{fig:ratio}b shows the factor NLO/LO for the quantity
defined as the ratio of the number of $W$ events to the number of $Z$
events at a given $X$.  As can be seen the NLO corrections to this
quantity are much less dependent on $X$ than the distribution
themselves.
Indeed, the corrections are similar for the $W$ and $Z$ mass-scaled
distributions and therefore cancel in the ratio.  This ratio could
then be used to measure the $W$ mass, with small theoretical
uncertainty.  Note that the measured mass and width of the Z are already 
used to calibrate the detectors~\cite{raja} in 
current analysis~\cite{wmassd0}.  
Compared to the standard transverse-mass method, the ratio
method will have a larger statistical uncertainty because it depends
on the $Z$ statistics, but a smaller systematic uncertainty because of
the use of the ratio.  This concept has now been verified in an experimental
analysis\cite{RATIO}.  Overall this ratio method might
therefore be competitive if the systematic uncertainty dominates
the overall uncertainty on $m_W$ in the transverse-mass method.  
It is beyond the
scope of this paper to study the sytematic uncertainties in detail;
in what follows we shall benchmark these uncertainties using the demonstated
CDF and D0 performance.  The ratio method can also be used with other
distributions, like the $p_T$-distribution of the charged lepton
itself, see~\cite{gandk}.

It is interesting to note that the average Bjorken-$x$ of the partons
producing the $W$ at LHC with the cuts considered in this paper is
$\sim 10^{-2}$ , compared to $\sim 10^{-1}$ at the Tevatron.  Without
the rapidity cut, the range of $x$ probed at the LHC is much larger,
going from below $10^{-3}$ to above $10^{-1}$.  The uncertainty due to
the parton distributions will thus be different at the LHC and
Tevatron.  Considering that this uncertainty might dominate in this
very high precision measurement, complementary measurements at the
Tevatron and LHC would be very valuable. It is not possible to
quantify this statement considering the present status of PDF
uncertainties~\cite{SnoPDF}.

At this time, it is obviously impossible to predict the overall
theoretical uncertainty at the LHC.  The present uncertainty of $\sim
30$~MeV from the $W$ production model\cite{wmassd0} would already
limit the precision of the mass measurement attainable in Run~II at
the Tevatron, so there is obviously great motivation to reduce such
uncertainties.  Part of our goal in writing this paper is to emphasize
that such motivation also exists for the LHC, by demonstrating its
potential for an extremely precise $W$ mass measurement.  In the rest
of this paper we shall assume that the theoretical uncertainty at the
LHC will be decreased to a value lower than the experimental
uncertainty.

\section{Experimental Uncertainties}

\begin{figure}[th]
\vskip 3 in
\includegraphics{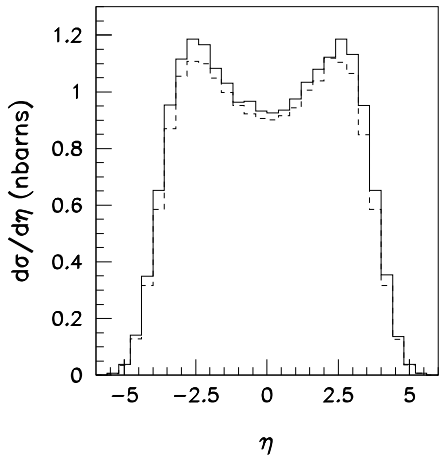}
\caption{Rapidity distribution of the charged lepton from single $W$
production at the LHC.
The histogram is for the NLO calculation and the dashed line for the 
LO.  See text for the cuts.}
\label{fig:eta}
\end{figure}

The single $W$ production cross section at the LHC, for charged lepton
$p_T > 20$~GeV/$c$ and pseudorapidity $|\eta| < 1.2$, and transverse
mass $65 GeV\leq m_T \leq 100GeV$, is about $4$ times larger than at
the Tevatron with the same 
cuts~\footnote{limiting the $W$ transverse momentum to be
less than 15 or 30~GeV does not significantly change this result.}.  
Scaling from the latest high-statistics $W$ mass measurement 
at D0~\cite{wmassd0}, where
$2.8\times 10^4$ $W \to e$ events were taken from an integrated
luminosity of 82~pb$^{-1}$, we then expect at the LHC $\sim 1.5 \times
10^7$ reconstructed $W\to e$ events in one year at low luminosity (for
$10fb^{-1}$).  Figure~\ref{fig:eta} shows that if the lepton rapidity
coverage at the LHC were increased above the $\pm 1.2$ assumed here, a
large gain in signal statistics would be obtained, since the rapidity
distribution is rather broad at the LHC energy.  The configuration for
which one Bjorken-$x$ is very large and the other one very small is
favored and creates the maxima at $|\eta | \sim 2.5$.
The gain would be of order two if leptons were accepted out to $|\eta|
< 2.5$, which is covered by the electromagnetic calorimetry of the
ATLAS~\cite{ATLAS} and CMS~\cite{CMS} experiments, and as high as a
factor of four for $|\eta| < 5$ which may be covered by other 
experiments~\cite{FELIX}.

As already noted, it is not straightforward to estimate the precision
with which $m_W$ can be determined because of the importance of
systematic effects; even a full GEANT simulation of a detector is
unlikely to include all of them.  We have therefore based our estimate
on a parametrization of the actual CDF and D0 $m_W$ uncertainties
developed in Ref.~\cite{tev2000} in order to extrapolate to higher
luminosity.  The parametrization includes the effect of the number of
interactions per crossing, $I_C$ (which degrades the missing $E_T$
resolution), and of those systematic effects which can be controlled
using other data samples (such as $Z$ bosons, $J/\psi$ mesons, etc.)
and which will therefore scale like $1/\sqrt{N}$.  This behavior
appears valid for the most important systematic uncertainties in the
present measurement, such as the energy scale determination,
underlying event effects, and the $p_T$ distribution of the $W$.  The
use of these parametrizations, of course, explicitly does not take
into account any of the detector improvements offered by the LHC
detectors over their Tevatron counterparts which were described
earlier.

The parametrized statistical and systematic uncertainties on $m_W$
are given by:
\begin{eqnarray}
\Delta m_W |_{stat}& = &12.1\,{\rm GeV} \sqrt{I_C\over N} \sim 4.4\,{\rm MeV}
\nonumber \\
\Delta m_W |_{sys}& = &17.9\,{\rm GeV} \sqrt{I_C\over N} \sim 6.5\,{\rm MeV}
\end{eqnarray}
where $N$ is the total number of events.  Taken at face value these
would suggest that $\Delta m_W \sim 8$~MeV could be reached. 
However, these parametrizations do not account for effects which do
not scale as $1/\sqrt{N}$.  Such systematic effects,
which are not yet important in present data, will probably limit the
attainable precision at the LHC.  There is however an opportunity 
to measure the $W$ mass to a precision of better than $\Delta m_W \sim
15$~MeV at the LHC.

It is worth noting that, while we have assumed that only one year of
operation at low luminosity is required to collect the dataset,
considerably longer would undoubtedly be required after the data are
collected in order to understand the detector at the level needed to
make such a precise measurement.

\section{Conclusions}

In conclusion, we see no serious problem with making a precise
measurement of $m_W$ at the LHC if the accelerator is operated at low
luminosity ($10^{33}\,{\rm cm}^{-2}\,{\rm s}^{-1}$) for at least a
year.  The cross section is large, triggering is possible, lepton
identification and measurement straightforward, and the missing
transverse energy should be well determined.  The QCD corrections to
the transverse mass distribution although larger than at the Tevatron,
still appear reasonable.  A precision better than $\Delta m_W \sim
15$~MeV could be reached, making this measurement the world's best
determination of the $W$ mass.  We feel that it is well worth
investigating this opportunity in more detail.

J.W. wishes to thank the Aspen Center for Physics
for its hospitality during the final stages of this work. 

%

\end{document}